\documentclass[prb,aps]{revtex4}
\usepackage{epsfig}
\usepackage{setspace}
\usepackage{indentfirst}
\usepackage{graphicx}
\usepackage[final]{pdfpages}
\begin{document}
%\draft
%\includepdf[lastpage=18,pages={1,{ },2-18}]{Transient_excitons}
%{\bf }

\begin{center}
\vskip 2cm
{\Large\textbf{Transient excitons at metal surfaces}} 
\vskip 1cm
\begin{center} 
{\large Xuefeng Cui$^{1}$, Cong Wang$^{1}$, Adam Argondizzo$^{1}$, Sean Garrett-Roe$^{2}$, Branko Gumhalter$^{3}$, \\
\vskip 3mm
and Hrvoje Petek$^{1}$}
\end{center}
\vskip 0.8cm
\begin{center}{ \large\it $^{1}$Department of Physics and Astronomy, University of Pittsburgh, Pittsburgh PA 15260 \\
\vskip 3mm
$^{2}$Department of Chemistry, University of Pittsburgh, Pittsburgh PA 15260 \\
\vskip 3mm
$^{3}$Institute of Physics, HR-10000 Zagreb, Croatia}
\end{center}
\end{center}
%\maketitle
\vskip 3cm

\begin{spacing}{2.0}
\noindent{\large\bf Abstract}

\large\hspace*{1cm}Excitons, electron-hole pairs bound by the Coulomb potential, are fundamental quasiparticles of coherent light-matter interaction energizing processes from photosynthesis to optoelectronics$^{1-5}$. Excitons are observed in semiconductors, and their existence is implicit in the quantum theory of metals, yet their appearance is tenuous due to the screening of the Coulomb interaction on few femtosecond timescale$^{6-8}$.Here we present direct evidence for the dominant transient excitonic response at a Ag(111) surface, which precedes the full screening of the Coulomb interaction, in the course of a three-photon photoemission process with $<$15 femtosecond laser pulses. Electron-hole pair interaction through the excitonic response introduces coherent quasiparticle correlations beyond the single-particle description of the optics of metals, which dominate the multi-photon photoemission process.

\large\hspace*{1cm}Reflection of light has made metal mirrors valued optical instruments since the bronze age$^{9}$. At the macroscopic level the coherent optical response of a metallic surface is well described by the classical Maxwell's equations. At the quantum level, a photon interacting with a metal surface polarizes an electron-hole \textit{(e-h)} pair to create an exciton-polariton, the quasiparticle of light-matter interaction1$^{10}$.The creation of excitons in insulators, semiconductors, and molecules provokes many-body, coherent optical processes, which have been studied in contexts of photosynthesis, vision, optical communication$^{1-5}$, etc. Yet in metals, the role of excitons remains uncharted, because screening building up on the timescale of plasma oscillation liberates bound states of the Coulomb potential$^{6-8}$. The dynamical response of a metal surface to light terminates either in reflection, which is coherent, or absorption, which can be detected through the photoelectric effect$^{11}$. 

\large\hspace*{1cm}We reveal the transient optical response of the Ag(111) surface triggered by electron from the Shockley surface state (SS) absorbing a photon to form instantaneously the primary exciton. In response, the screening charge density fluctuations cause the bare Coulomb potential binding the exciton to wane, and simultaneously the image potential (IP) binding the electron to its screening image charge, to emerge as the asymptotic state of the quasiparticle interacting with its medium. This transient regime of the coherent excitonic polarization is revealed by a new mode of surface photoemission via m$^{th}$-order multiphoton absorption. 

\large\hspace*{1cm}Coherent optical response of metals has been studied by energy (E) and momentum (k) resolved multiphoton photoemission (mPP) spectroscopy. Interferometric time-resolved 2PP measurements12 determined surface and bulk electron and hole dephasing on femtosecond timescales$^{13-15}$. 2PP studies on silver have probed the dephasing and lifetimes of intermediate IP states$^{16-18}$. Attosecond studies have revealed band-dependent photoelectron emission delays$^{19}$. In every case, light was thought to excite transitions between preexisting electronic bands.

\large\hspace*{1cm}Excitons in metals have been more illusive. The excitonic response has been discussed in the context of theoretical modeling of the dielectric functions of metals$^{20-22}$. The transient excitonic response, however, has been proposed to have potentially observable consequences in ultrafast mPP spectroscopy$^{23,24}$. 

\large\hspace*{1cm}Here we reveal the excitonic response in mPP spectra of the Ag(111) surface upon excitation near the two-photon IP SS transition (Fig. 1) by intense, broadly tunable laser pulses with $<$15 fs duration. For non-resonant excitation, the SS and IP state energy-momentum distributions [$E(k)$: Figs. 2a, h] appear as dispersive bands such as reported in 2PP studies employing two-colour incoherent excitation$^{18}$. A dramatic change in mPP spectra arises when the laser spectrum overlaps with the two-photon IP SS transition at $\hbar\omega_{res} =1.93\pm0.02 eV$. A new feature that dominates the 3PP spectra (Figs. 2b-g), but does not exist in the single-particle band structure (Fig. 1a), appears at the final state energy $E_f=5.82\pm0.03 eV$ above the Fermi level ($E_F$). Because of its nondispersive character, more than 100 times higher intensity than the non-resonantly excited SS and IP bands (Figs. 2a, h), and photon energy independent photoemission energy, we attribute it to the transient exciton (TE).

\large\hspace*{1cm}Plotting $E_f$ for the SS, IP, and TE features against the excitation photon energy,  , we find linear behavior with approximate slopes of three, one, and zero (Fig. 3). This would be expected within Einstein's model of the photoelectric effect$^{11}$, where the single-particle energy level and define the photoelectron energy, if SS, IP, and TE were the initial, penultimate, and final states in a 3PP process. Because a discrete state at $E_f=5.82 eV$ that could explain TE does not exist (Fig. 1a)$^{25}$, we will argue that TE manifests a correlated mode of non-Einsteinian photoemission.

\large\hspace*{1cm}Gumhalter \textit{et al}. elaborated an excitonic model for 2PP from the occupied Shockley surface states of Ag(111) or Cu(111) surfaces via the intermediate IP states, which approximately describes our experiment$^{24}$. Whereas the SSs are Bloch states derived by the surface boundary conditions, the IP states emerge through the many-body screening response26. Excitation of SS within the Drude free-electron continuum can only create the primary exciton state$^{24}$, because there is no transition moment to the unformed IP state. In response to excitation, charge density fluctuations screen the Coulomb field and thereby produce the screened electron and hole quasiparticles. Upon saturation of the screening, the IP state emerges as on-the-energy-shell, asymptotically evolved excited surface electron state$^{24}$. The saturation, however, is metal dependent: in silver it takes $\sim$15 fs because of its sharp and low frequency (3.7 eV) surface plasmon, whereas for copper it is complete within $\sim$2 fs$^{24}$. 

\large\hspace*{1cm}The surface exciton on Ag(111) is composed of the SS hole interacting with the upper $sp$-band ($U_{sp}$) electron Bloch states (Fig. 1a)$^{24,27}$. Wavefunction of the exciton described by N, the quantum number for relative motion, and \textbf{K} the total center-of-mass momentum wavevector, may be written as a superposition (wavepacket) of the coupled hole $\bf{k}_h$ and electron $\bf{k}_e$ momentum states$^{27}$, 
$$\mid\bf{K},N\rangle = \sum_{\bf{k}_e,\bf{k}_h}\Psi_{\bf{k}_e,\bf{k}_h}^{\bf{K},N}\mid\bf{k}_e,\bf{k}_h\rangle\eqno{(1)}$$
\large where $\Psi_{\bf{k}_e,\bf{k}_h}^{\bf{K},N}$ is the amplitude of the constituent band states contributing to the exciton in the $k$-space, and $\bf{K}\approx0$ is imparted by the photon$^{27}$. The excitonic eigenenergies (Fig. 1b) are obtained by solving the Schr$\ddot{o}$dinger equation for the bare Coulomb potential of the SS-hole charge density$^{24}$. 

\large\hspace*{1cm}The salient features of an exciton expressed in Equation (1) directly explain two facets of the TE behavior: $i$) the nondispersive TE photoemission derives from it being a localized superposition state in the relative coordinate space constrained by $\bf{K}=\bf{k}_e+\bf{k}_e\approx0$; and $ii$) its optical transition moment enhancement derives from the summation over the available interband transitions. The congruence of the $k_{||}$ ranges of TE and SS attests to TE being the superposition of all SS states within its 63 meV occupied bandwidth.

\large\hspace*{1cm}To illuminate the role of TE as a precursor to IP in a two-photon IP SS transition, we image the photoelectron E($k$) distributions as a function of delay $\tau$ between identical, collinear, phase-correlated pump-probe pulses. Acquiring E($k$) distributions in delay intervals of $\tau\sim100$ attoseconds for $\hbar\omega_{laser}=2.05 eV$ records a movie of the TE dynamics (Supplementary Movie S1). The interferogram for $k_{||}=0 \AA^{-1}$ and E($k$) distribution for $\tau=0$ fs (Fig. 4a and b) are cross-sections through the three-dimensional ($E,k,\tau$) data. Correlation traces (Fig. 4c-e) representing cross-sections through the interferogram for $E_f$ marked by the lines cutting Fig. 4a reveal the E($k$)-resolved coherent polarization dynamics of SS, IP, and TE.

\large\hspace*{1cm}Instead of single-point correlations, we analyze the co-relationship between the SS, IP, and TE polarizations by Fourier transforming the interferograms for three representative values of $k_{||}$ to obtain 2D spectra of the linear polarization $vs.$ photoelectron energy relative to the SS band minimum (Figs. 4f-h; Supplementary Fig. S1 gives the complete 2D spectra including the weak nonlinear components). We note that even for a detuning $\Delta=\omega_{laser}-\omega_{res}=0.12\ eV$ the TE response at the polarization energy $\hbar\omega_{res}$ dominates the 2D spectra above the IP state at $\hbar\omega_{laser}$ (Figs. 4f, g).  This indicates that the internal field at the IP $\gets$ SS two-photon resonance drives the surface response more effectively than the external laser field. Furthermore, the disposition of the dominant TE and weaker SS responses along the line with the slope of 1/3 (Figs. 4f, g) signifies their origin in coherent 3PP processes of primary states driven by the linear polarization fields.

\large\hspace*{1cm}Being retarded the IP state cannot participate in a coherent 3PP process, and therefore its disposition within the 2D spectra is distinct from those of TE and SS (Figs. 4f-h). Because the IP state appearance does not correlate with the occupied $k_{||}$ range of SS (Fig. 4h), and it alignment is with the slope of $\frac{1}{2}$, we conclude that it is excited from the lower $sp$-band ($L_{sp}$: Fig. 1a) by absorption of two photons at $\hbar\omega_{laser}$ from the external field. The weak dependence of the IP state with $k_{||}$ indicates that for $\Delta=0.12\ eV$ the $L_{sp}$ channel dominates the 3PP via the IP state, even when the TE channel is available. The emergence of the IP state from the TE channel and its subsequent photoemission by a single photon process, however, can be discerned in the additional density in 2D spectra for $\mid k_{||}\mid<0.07\ \AA^{-1}$ which extends between TE and IP features along the line with a slope of 1 (Figs. 4f, g). The presence of two channels for IP state photoemission via interactions with the $\hbar\omega_{res}$ and $\hbar\omega_{laser}$ fields leads to the polarization beating in its correlation trace at $\tau\cong20\ fs$ (Fig. 4d). A frequency analysis of the interferogram in Fig. 4a shows that the IP state photoemission driven by $\hbar\omega_{laser}$ dominates for $<20\ fs$ to be overtaken by the more slowly dephasing TE channel at $\hbar\omega_{res}$. As $\Delta\to0$, model simulations of two-pulse correlations (Supplementary Fig. S2) together with 3PP measurements (Fig. 2) support the scenario for the dominant IP state creation from the excitonic manifold.

\large\hspace*{1cm}The 2D spectra explain the final characteristic of the excitonic photoemission, namely the $\hbar\omega_{laser}$ independent TE photoemission energy. Upon femtosecond pulse illumination, an electron is excited by two-photon interaction from SS through a manifold of excitonic states converging in a quasicontinuum to the bottom of $U_{sp}$ (Fig. 1b). Concurrently, the screening response involving off-the-energy-shell transients dominated by the virtual surface plasmon excitation and constrained by the Heisenberg energy-time uncertainty evolves the bare exciton into its saturated form, the $n=1$ IP electron and the corresponding SS hole quasiparticle pair$^{28}$. In the course of this quantum kinetic evolution the virtual excitations die off through destructive interference and only the energy conserving excitations persist. Therefore, a local field can grow only at $\hbar\omega_{res}$ associated with the coherent coupling of the entire SS-band to form the asymptotic IP state. Because TE photoelectrons emerge by multiple quantum absorption from the excitonically enhanced local field at  $\hbar\omega_{res}$ , the TE energy does not track $\hbar\omega_{laser}$$^{29,30}$.

\large\hspace*{1cm} The correlated TE photoemission and its associated local field are manifestations of the fundamental screening response of solid-state matter to polarization by external optical fields. Although excitons are detected as stable quasiparticles in semiconductors and insulators, the dynamical dielectric response involved in their formation and propagation is universal, and should be observable in other materials. For example, the new mode of local field-induced photoemission driven by multiple-quantum transient exciton interactions in silver may illuminate the multi-exciton generation in organic films1. The inducing and probing of strong $e-h$ correlations even in a noble metal silver by coherent multiphoton photoemission spectroscopy, holds promise for extending to nonequilibrium quasiparticle dynamics in more strongly correlated materials. \\

\noindent{\large\bf Methods summary}

\setlength{\parindent}{0pt}\large\textbf{Femtosecond laser excitation. }The photoexcitation source for the mPP measurements is a noncollinear parameter amplifier (NOPA) system pumped by Clark MXR Impulse fiber-laser oscillator-amplifier system. The NOPA system is used $>$100 mW average power at 1.25 MHz repetition rate with $<$15 fs pulse duration. The p-polarized light incident at 45$^{\circ}$ from the surface normal is focused onto $\sim100 \mu m$ diameter spot at the Ag(111) sample. 

\setlength{\parindent}{0pt}\large\textbf{Interferometric pump-probe delay scanning. }Identical pump-probe pulse replicas are generated in a self-made Mach-Zehnder Interferometer (MZI)$^{12,13}$. The pump-probe delay is scanned with a piezoelectrically actuated translation stage at a rate of 7.9 fs/s. Recording interference fringes at the center laser wavelength by passing the secondary output of the MZI through a monochromator, and recording the resulting interferogram with a photodiode calibrates the delay scanning. After data acquisition, the about 200 interferometric scans are combined in software using the calibration interference fringes.

\setlength{\parindent}{0pt}\large\textbf{Photoelectron imaging. }The mPP photoelectron images are recorded with a Specs Phoibos 100 electron spectrometer equipped with a 3D-DLD delay-line photoelectron counting detector. For each interferometer pump-probe delay scan, 4096 E($k$) images are taken with an integration time of 12ms/image.

\setlength{\parindent}{0pt}\large\textbf{Sample. }The single crystal Ag(111) surface is prepared by conventional surface science methods in an ultrahigh vacuum chamber with a base pressure of $<10^{-10}$ mbar. During the measurements the sample is cooled to $\sim$100 K.

\setlength{\parindent}{0pt}\large\textbf{Figures}

\begin{figure}[htbp] 
\centering\includegraphics[width=6.5in]{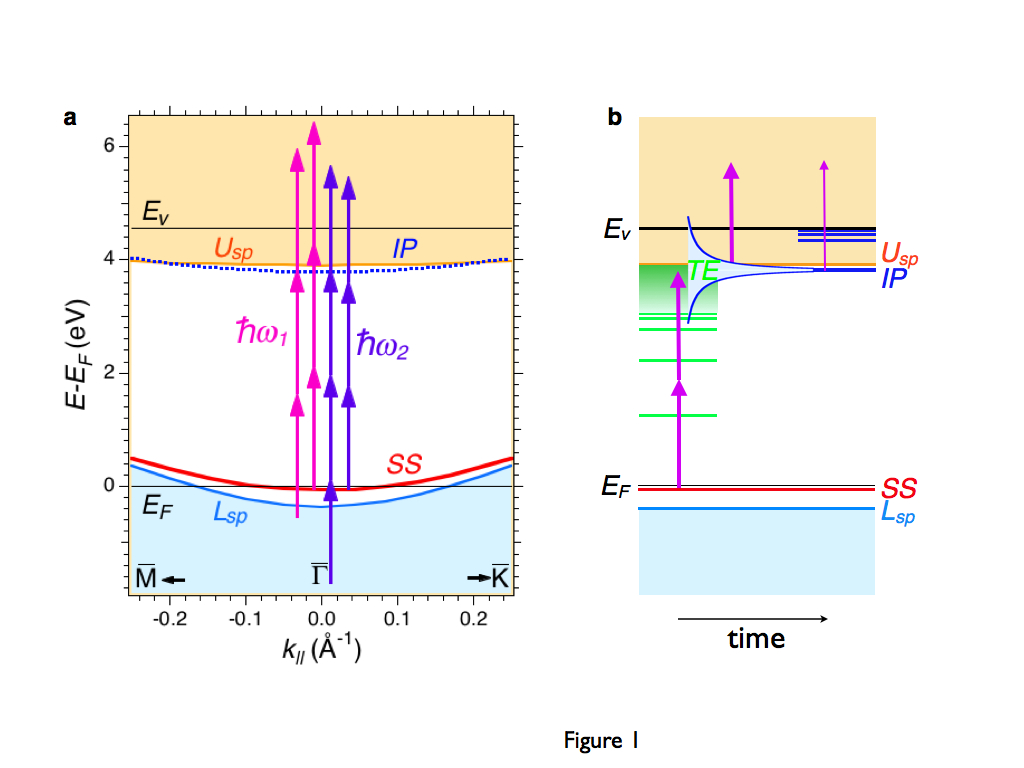}  
\end{figure} 
\setlength{\parindent}{0pt}\large\textbf{Figure 1. The electronic band structure and multiphoton photoemission processes at Ag(111) surface. a}, The surface projected band structure of Ag(111) has a band gap extending between -0.4 and 3.9 eV from the lower to the upper $sp$-band (L$_{sp}$ and U$_{sp}$). Within the band gap, the Shockley surface state, SS (red line) and $n=1$ IP state (blue dotted line) with minima at -0.063 and 3.79 eV form quantum wells at the metal-vacuum interface. SS is occupied to $\mid k_{||}\mid=0.07\ \AA^{-1}$ where it intersects the Fermi level ($E_F$). The $n=1$ IP state is the first member of a Rydberg-like series converging to the vacuum level, $E_v$$^{18,26}$. The vertical arrows indicate independent excitation pathways for 3PP and 4PP via the initial SS or the penultimate IP states for $\hbar\omega_1=2.20\ eV$ and $\hbar\omega_2=2.20\ eV$. \textbf{b}, The excitation scheme of IP states via the transient exciton, TE, manifold. The eigenstates of the excitonic manifold are obtained by solving the Schr$\ddot{o}$dinger equation for bare Coulomb potential of the SS hole (green lines); they converge in a quasi-continuum (shaded green) to the bottom of U$_{sp}$$^{24}$. Charge density fluctuations screening the Coulomb field evolve the bare exciton into the fully screened IP state. The coherent 3PP measurements follow the time evolution of the polarization amplitude within the excitonic manifold towards the asymptotic IP state under the influence of time-dependent potentials and energy-time uncertainty, which is implied by the time-evolving width of the IP state. 
\newpage
\begin{figure}[htbp] 
\centering\includegraphics[width=6.5in]{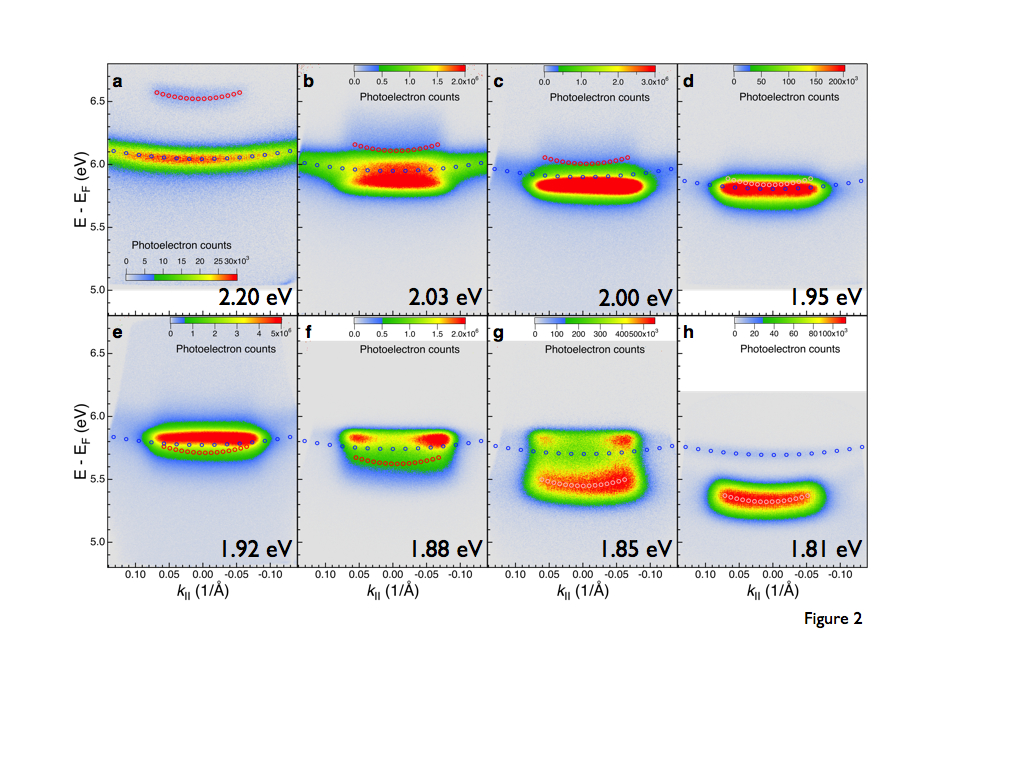}  
\end{figure} 
\setlength{\parindent}{0pt}\large\textbf{Figure 2. 3PP and 4PP E($k$) distributions for photon excitation energies near the two-photon IP SS resonance. a} and \textbf{h}, For nonresonant excitation at $\hbar\omega_{laser}$=1.81 and 2.20 eV. For nonresonant excitation at  1.81 and 2.20 eV the SS and IP bands appear with E($k$) dispersions consistent with the band structure and excitation processes indicated in Fig. 1a; the red and blue circles denote their expected parabolic dispersions. \textbf{b-g}, Upon tuning $\hbar\omega_{laser}$ into the two-photon $IP\gets SS$ resonance, a new feature characterized by enhancement of the transition moment, non-dispersive, i.e. flat, E(k) distribution spanning the occupied $k_{||}$ range of SS, and photon energy independent photoelectron energy appears at $E_f=5.82\pm0.03eV$. We attribute these characteristics to a transient excitonic state created by excitation of an electron from SS via the excitonic manifold, through a two-photon resonance involving the energy conserving IP state. The photoelectron energies are given with respect to $E_F$.

\begin{figure}[htbp] 
\centering\includegraphics[width=6.5in]{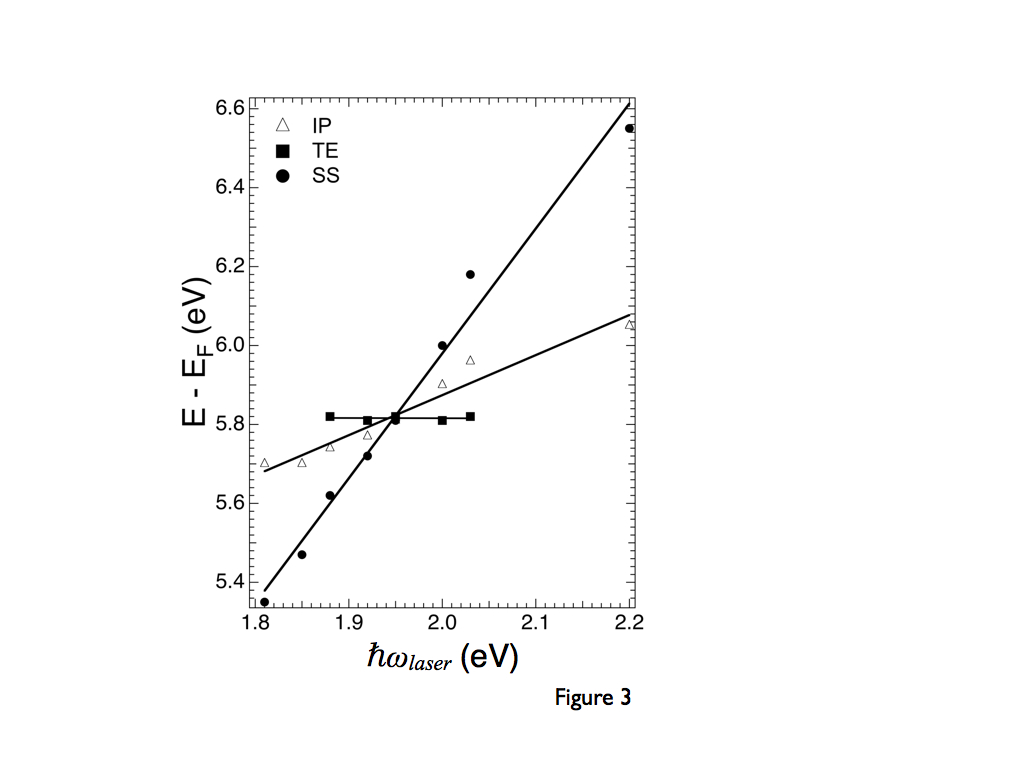}  
\end{figure} 
\setlength{\parindent}{0pt}\large\textbf{Figure 3. The tuning of the spectroscopic features with photon energy. }The photoelectron energies of SS, IP, and TE states in Fig. 2 are plotted vs. the excitation photon energy. According to Einstein's model, the photoelectron energy in a photoemission process is defined by the photon energy. In a multiphoton process, the tuning of the photoelectron energy $E_f$ of a spectroscopic feature is determined by the number of photons times the photon energy required to excite an electron from it to above $E_v$. Therefore, in the 3PP process, the initial state (SS) is expected to tune with a slope of three and the penultimate state (IP) with a slope of one, as observed. Accordingly, the slope of zero would attribute the TE feature to a sharply defined state at $E_f=5.82\pm0.03eV$, which does not exist in the band structure of Ag. The TE behavior therefore implies a previously unknown, non-Einsteinian photoemission process where resonantly enhanced local field associated with the excitonic polarization of the sample excites the TE photoemission. 

\begin{figure}[htbp] 
\centering\includegraphics[width=6.5in]{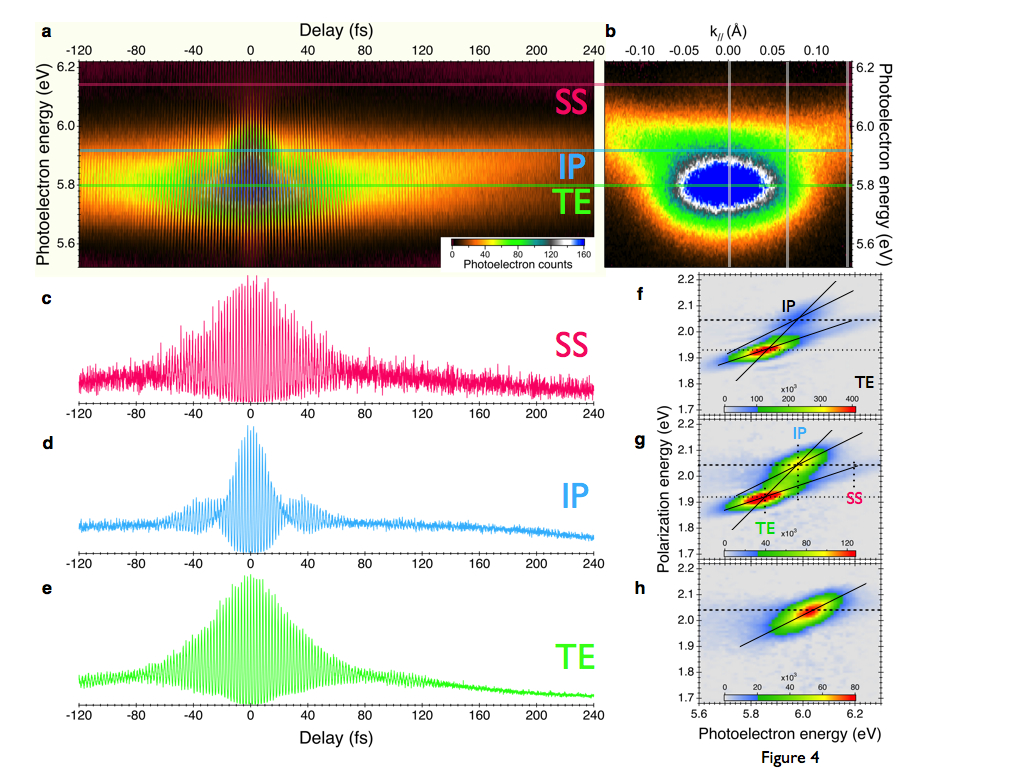}  
\end{figure} 
\setlength{\parindent}{0pt}\large\textbf{Figure 4.  3D photoemission spectra for near-resonant two-photon $IP\gets SS$ excitation. a}, The interferogram of photoelectron counts vs. photoelectron energy and time delay between interferometrically scanned pump-probe pulses for $k_{||}=0$. The $\hbar\omega_{laser}$ = 2.05 eV is detuned from the resonance energy, $\hbar\omega_{res}$ = 1.93 eV. \textbf{b}, The E($k$) image at zero pump-probe delay. The horizontal lines indicate the energies of the SS, IP, and TE signals, and the vertical lines indicate momenta for the 2D spectra in \textbf{f-h}. \textbf{c-e} Correlation measurements 13 obtained by taking cross sections through the interferogram in a at the energies of SS, IP, and TE. The oscillations with approximately the period of the carrier wave (~2 fs) represent the interference between the pump and probe pulse-induced polarizations excited in the sample. \textbf{f-h}, 2D photoelectron spectra obtained by Fourier transforming the interferometric scans such as in \textbf{a} for $k_{||}=0.0$ (\textbf{f}), 0.067 (\textbf{g}) and 0.134 $\AA^{-1}$ (\textbf{h}). The Fourier axis is labeled the ?polarization energy?. Higher order polarization components are substantially weaker (Supplementary Fig. S1). The 2D plots show the correlation between the coherent polarization and photoemission energy. The dashed and dotted lines indicate $\hbar\omega_{laser}$ and $\hbar\omega_{res}$, the vertical lines indicate the energies of the SS, IP, and TE signals, and the full lines designate the slopes of $\frac{1}{3}$, $\frac{1}{2}$, and 1.

\end{spacing}

\newpage
\noindent{\large\bf References}
\begin{enumerate}
\item Chan, W.-L. et al. Observing the Multiexciton State in Singlet Fission and Ensuing Ultrafast Multielectron Transfer. \textit{Science} 334, 1541-1545 (2011).
\item Gibbs, H.M., Khitrova, G., $\&$ Koch, S.W. Exciton-polariton light-semiconductor coupling effects. \textit{Nat Photon} 5, 273-273 (2011).
\item Turner, D.B. $\&$ Nelson, K.A. Coherent measurements of high-order electronic correlations in quantum wells. \textit{Nature} 466, 1089-1092 (2010).
\item Cundiff, S.T. $\&$ Mukamel, S. Optical multidimensional coherent spectroscopy. \textit{Physics Today} 66, 44-49 (2013).
\item Lee, H., Cheng, Y.-C., $\&$ Fleming, G.R. Coherence Dynamics in Photosynthesis: Protein Protection of Excitonic Coherence. \textit{Science} 316, 1462-1465 (2007).
\item Edwards, P.P., Lodge, M.T.J., Hensel, F., $\&$ Redmer, R. '... a metal conducts and a non-metal doesn't'. \textit{Phil. Trans. R. Soc. }A 368, 941-965 (2010).
\item Silkin, V.M., Kazansky, A.K., Chulkov, E.V., $\&$ Echenique, P.M. Time-dependent screening of a point charge at a metal surface. \textit{J. Physics: Condens. Matter} 22, 304013 (2010).
\item Huber, R. et al. How many-particle interactions develop after ultrafast excitation of an electron-hole plasma. \textit{Nature} 414, 286-289 (2001).
\item Enoch, J.M. History of Mirrors Dating Back 8000 Years. \textit{Optometry \& Vision Science} 83, 775-781 (2006).
\item Hopfield, J.J. Theory of the Contribution of Excitons to the Complex Dielectric Constant of Crystals. \textit{Phys. Rev. }112, 1555 (1958).
\item Einstein, A. Über einen die Erzeugung und Umwandlung des Lichtes betreffenden heuristischen Standpunkt. \textit{Annalen der Physik} 17, 132 (1905).
\item Petek, H. $\&$ Ogawa, S. Femtosecond Time-Resolved Two-Photon Photoemission Studies of Electron Dynamics in Metals. \textit{Prog. Surf. Sci.} 56, 239-310 (1997).
\item Ogawa, S., Nagano, H., Petek, H., $\&$ Heberle, A.P. Optical dephasing in Cu(111) measured by interferometric two-photon time-resolved photoemission. \textit{Phys. Rev. Lett}. 78, 1339-1342 (1997).
\item Petek, H., Nagano, H., $\&$ Ogawa, S. Hole Decoherence of d Bands in Copper. \textit{Phys. Rev. Lett.} 83, 832-835 (1999).
\item Güdde, J. et al. Time-Resolved Investigation of Coherently Controlled Electric Currents at a Metal Surface. \textit{Science} 318, 1287-1291 (2007).
\item Giesen, K. et al. Image Potential States Seen via Two-Photon Photoemission and Second Harmonic Generation.\textit{ Physica Scripta} 35, 578 (1987).
\item Schoenlein, R.W., Fujimoto, J.G., Eesley, G.L., $\&$ Capehart, T.W. Femtosecond relaxation dynamics of image-potential states. \textit{Phys. Rev. B} 43, 4688-4698 (1991).
\item Marks, M. et al. Quantum-beat spectroscopy of image-potential resonances. \textit{Phys. Rev. B }84, 245402 (2011).
\item Cavalieri, A.L. et al. Attosecond spectroscopy in condensed matter. \textit{Nature} 449, 1029-1032 (2007).
\item Mueller, F.M. $\&$ Phillips, J.C. Electronic Spectrum of Crystalline Copper. \textit{Phys. Rev}. 157, 600 (1967).
\item Fong, C.Y. et al. Wavelength Modulation Spectrum of Copper. \textit{Phys. Rev. Lett.} 25, 1486 (1970).
\item Marini, A. $\&$ Del Sole, R. Dynamical Excitonic Effects in Metals and Semiconductors. \textit{Phys. Rev. Lett.} 91, 176402 (2003).
\item Schöne, W.-D. $\&$ Ekardt, W. Transient excitonic states in noble metals and Al. \textit{Phys. Rev. B} 65, 113112 (2002).
\item Gumhalter, B., Lazi$\acute{c}$, P., $\&$ Do$\breve{s}$li$\acute{c}$, N. Excitonic precursor states in ultrafast pump?probe spectroscopies of surface bands. \textit{phys. stat. sol.} (b) 247, 1907-1919 (2010).
\item Miller, T., Hansen, E.D., McMahon, W.E., $\&$ Chiang, T.C. Direct transitions, indirect transitions, and surface photoemission in the prototypical system Ag(111). \textit{Surf. Sci.} 376, 32-42 (1997).
\item Echenique, P.M. $\&$ Pendry, J.B. Theory of image states at metal surfaces. \textit{Prog. Surf. Sci.} 32, 111 (1990).
\item Elliott, R.J. Intensity of Optical Absorption by Excitons. \textit{Phys. Rev. }108, 1384-1389 (1957).
\item Gumhalter, B. Stages of hot electron dynamics in multiexcitation processes at surfaces: General properties and benchmark examples. \textit{Prog. Surf. Sci.} 87, 163-188 (2012).
\item Merschdorf, M., Kennerknecht, C., $\&$ Pfeiffer, W. Collective and single-particle dynamics in time-resolved two-photon photoemission. \textit{Phys. Rev. B }70, 193401 (2004).
\item Winkelmann, A. et al. Resonant coherent three-photon photoemission from Cu(001). \textit{Phys. Rev. B} 80, 155128-155129 (2009).
\end{enumerate}

\newpage
\vskip 3 cm

\begin{center} {\Large Supplementary Information for} 
\end{center}
\vskip 4cm
\begin{center} {\huge{\bf Transient Excitons at Metal Surfaces}}
\end{center}
\vskip 1cm
\begin{center} {\Large
Xuefeng Cui, Cong Wang, Adam Argondizzo, Sean Garrett-Roe,\\
\vskip 3mm
 Branko Gumhalter, Hrvoje Petek$^{*}$}
\end{center}
\vskip 4mm
\begin{center}{\small{ \it $^{*}$Corresponding author, E-mail: petek@pitt.edu (HP)}}
\end{center}
%\date{\today}
%\begin{abstract}
%\end{abstract}
%\maketitle

\vskip 2 cm

{\large 
 {\bf This file includes:}

\vskip 5 mm
\begin{enumerate}

\item Supplementary Movie S1

\item Supplementary Data:  2D photoelectron spectra for different photoelectron momenta \&  Fig. S1 

\item Supplementary Simulation: Interferometric 3PP spectra \& Fig. S2

\item Supplementary References R1-R12
\end{enumerate}
}
\newpage

\newcommand{\bq}{\begin{equation}}
\newcommand{\eq}{\end{equation}}
\newcommand{\barr}{\begin{eqnarray}}
\newcommand{\earr}{\end{eqnarray}}

\noindent{\bf 1. Supplementary Movie S1: 3D photoelectron imaging of the TE dynamics}

A movie of the $E(k)$ distributions of three-photon photoemission (3PP) from Ag(111) surface with 2.05 eV photon excitation taken at pump-probe delay intervals of $\sim$100 as.  The peak distributions of the TE, IP and SS features determined with single pulse excitation are indicated by the green, blue and red shading.  At finite delays the interference between the pump and probe pulse excited polarizations causes the distributions to acquire structure and shift in energy from their zero delay positions.

\vskip 3 cm

\begin{center}
\rotatebox{0}{\epsfxsize=7.8cm \epsffile{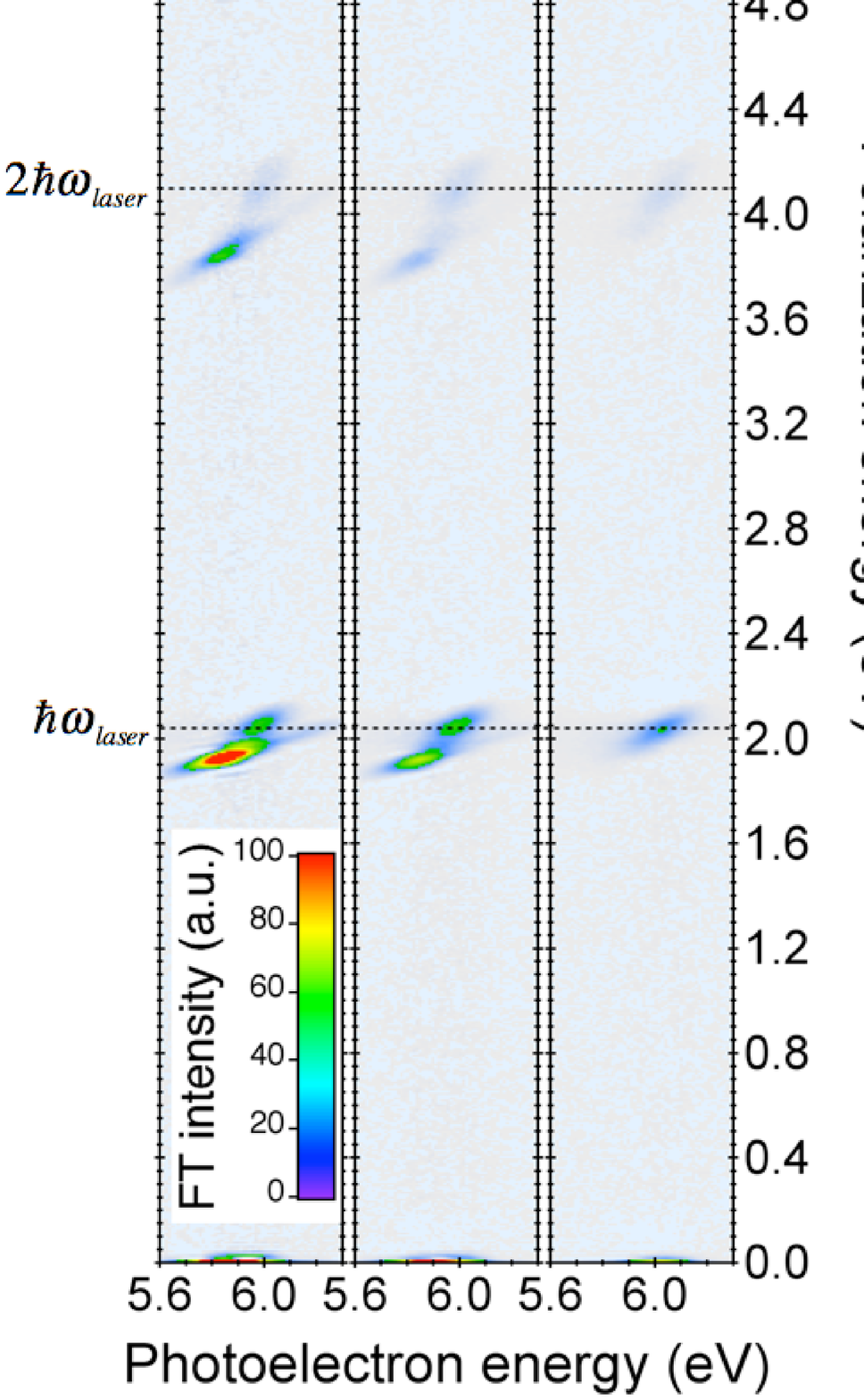} } %TE_Science_FigS1
\end{center}
\noindent {\small {\bf Figure S1.}: 2D photoelectron spectra for selected photoelectron parallel momenta obtained with  $\hbar\omega_{laser}=2.05$ eV.(a-c) give the 2D spectra for $k_{\parallel}$=0, 0.07 and 0.14 \AA$^{-1}$. The dashed lines indicate the polarization at  $\hbar\omega_{laser}=2.05$ eV and its second and third harmonics.}

\vskip 5mm

\noindent{\bf 2. Supplementary data: 2D photoelectron spectra for different photoelectron momenta}

The Fourier transform (FT) of the 3D photoelectron imaging data with respect to the delay time $\tau$ obtains 2D photoelectron spectra for different photoelectron emission angles (Figure S1). 
The ordinate is the energy of the coherent polarization components contributing to the 3PP process in eV, as deduced by FT of interferograms extracted from the Supplementary Movie S1 for $k_{\parallel}=0$ (Fig. 4a), 0.07 and 0.14 \AA$^{-1}$. The abscissa is the photoelectron energy with respect to the bottom of the SS band, which is the reference energy for the TE excitation. The SS band minimum is 0.063 eV below $E_{F}$, and therefore this energy scale is nearly the same as the photoelectron energy relative to $E_{F}$ referenced elsewhere in the publication.

The 2D spectra show individual components of the induced linear and nonlinear polarization as elliptical features. The major axis of these features represents the bandwidth of the excitation pulse that is used to drive the coherent response. The minor axes represent the frequency bandwidth implied by the free-induction decay of each component of the coherent response, which is in general slower than the excitation pulse duration. The inclination of each ellipse is determined by the ratio of the order of the coherent polarization to the order of the coherent response leading to the 3PP process. For example, the TE and SS features driven by the linear polarization appear through a third-order coherent process, hence the slope of the major axis of their ellipses is $\frac{1}{3}$.

The spectra in Fig. S1 show that although the coherent polarization has contributions at the second and third harmonics of $\hbar\omega_{laser}$, the dominant response is at the resonant frequency $(E_{IP}-E_{SS})/2= \hbar\omega_{res} = 1.93$ eV which is within the intraband, Drude absorption, continuum of silver. Here $\omega_{laser}$ is the carrier wave frequency of the laser pulses and $E_{SS}$ and $E_{IP}$ are the electron energies at the bottoms of SS-band and relaxed first IP-band, respectively. Weaker responses at the harmonics of $\hbar\omega_{res}$ can also be seen to dominate the harmonics of $\hbar\omega_{laser}$. Contrary to the expectation of an instantaneous dephasing within the continuum, the TE coherent polarization persists on $>$50 fs time scale according to the interferogram in Fig. 4e. The simultaneous measurement of the much faster dephasing dynamics for the IP-state is one indication that the coherent response reflects the properties of Ag(111) surface. The 2D spectra are presented on the same  intensity scale in order to facilitate the comparison of the various orders and components of the coherent polarization for different photoemission momenta.

\vskip 5 mm

\noindent{\bf 3. Supplementary simulation: Excitonic features in interferometric 3PP from Ag(111) surface state band}

The most striking results of the measurements described in the main text are Figs. 2c-e and 4e. Figures 2c-e show the 3PP spectra dominated by a nondispersive feature which exhibits the photon energy independent photoelectron energy at $5.82\pm 0.03$ eV and spans the occupied $k_{\parallel}$ range of the SS-band on Ag(111) surface. Following the energy level diagram in Fig. 1b, these characteristics were attributed to a transient exciton (TE) state created by electron excitation from SS-state via the excitonic manifold, through a two-photon resonance involving the relaxed IP-state.  Figure 4e shows the two-pulse correlation measurement of 3PP from the SS-band  excited with $\hbar\omega_{laser}=2.05$ eV, which is recorded at the TE photoemission energy and designated TE. Prominent features of the TE two-pulse corelation measurement are its extension over a  longer delay interval than for the IP-state displayed in Fig. 4d, and oscillation at $\hbar\omega_{res} = 1.93$ eV.   In accordance with the assignments of the spectra of Figs. 2c-e, this feature was interpreted as yet another manifestation of the manifold of excitonic states which is instantly available to two-photon induced polarization and evolves in the course of screening of the excited electron-hole (e-h) pair. Here we shall demonstrate that, in addition to the calculations of primary excitonic levels yielding the TE energetics shown in Fig. 1b, also model simulations of two-pulse correlations of the 3PP process support the occurrence of transient excitonic states, which in the interplay with the emergent IP-states give rise to the coherent polarization fringes that extend significantly  beyond  the cross-correlation of the applied pulses. 

The demonstration starts from the  perturbative picture of 3PP in which the absorption of the first photon excites electron from an SS-band state $|\phi_{SS}\rangle$ to a virtual state in the surface projected {\it s,p}-band gap, where it is instantly subject to the bare (unscreened) Coulomb potential of the SS-hole. The latter interaction supports a manifold of primary excitonic bound states$^{24}$ labeled by the quantum number $N$. The manifold consists of the low lying discrete states $|\phi_{N}\rangle$ for $N=1,2,3, \ldots$ etc. and the high excited Rydberg-like or distorted Kepler-like states $|\phi_{Kep}\rangle$ for $N\rightarrow \infty$, which make a quasicontinuum below the upper edge of the band gap (see Fig. 1b of the main text). Subsequent onset of dynamical screening of the excited e-h pair has a twofold effect on the potentials governing the dynamics and energetics of the pair. First, the initially dominating monopole component of the primary excitonic potential is efficiently screened by the potential of developing hole image charge and thereby strongly reduced to a much weaker dipolar interaction. Second, the simultaneous build up of the electron image charge gives rise to the surface image potential acting on the excited electron. Both processes proceed at the same pace on the time scale set by the saturation of surface screening charge, i.e. are governed by the transient factor discussed and illustrated in Sec. 5 and Fig. 5 of Ref. 24, respectively. The resulting effective time-dependent potentials $V^{exc}(t)$ and $V^{IP}(t)$ governing the excited electron motion relative to the SS-hole and the surface, respectively, support the bound states which evolve in the course of screening from the primary excitonic states deriving from the initial unscreened excitonic potential $V^{exc}(t=0)$ into the relaxed IP-states deriving from the electron image potential $V^{IP}(t=t_{sat})$ where $t_{sat}\sim 15$ fs is the screening saturation time (cf. Fig. 5 of Ref. 24). The associated bound state energy spectrum transforms from the primary excitonic spectrum spanning a large part of the {\it s,p}-band gap into a much narrower spectrum extending from the lowest IP-level to the vacuum level $E_{V}$. Electron propagation from the initial set of bound states over into the set of relaxing IP-states is a highly nonadiabatic process described by a complicated wavefunction $\psi(t)$ whose time dependence is driven by $V(t)=V^{exc}(t)+V^{IP}(t)$ and  whose solution requires the diagonalization of the full Hamiltonian $H(t)=H_{0}+V(t)$ at each instant of time$^{R1,R2}$ (for example of $V(t)$ see Fig. 6 in Ref. 24). Instead of attempting this formidable procedure we shall represent $\psi(t)$ as a linear combination of the wavefunctions $\phi_{exc}(t)$ and $\phi_{IP}(t)$ diagonalizing $H^{exc}(t)=H_{0}+V^{exc}(t)$ and $H^{IP}(t)=H_{0}+V^{IP}(t)$, respectively, and in the following consider the time evolution of $\psi(t)$ as a sequence of transitions between these two sets of diabatic states.

Building on the afore introduced scenario we assert that in the photon energy interval in the experiments described in the main text the first step of perturbative photoemission from SS-band is induced by one photon absorption from either pulse and proceeds offresonantly via the instantly available lowest primary excitonic state$^{24}$ $|\phi_{N=1}\rangle$ (or few lowest states since the other instantly available non-exciton states lie in the band above the surface projected {\it s,p}-band gap and hence are much more detuned from the pulse photon frequency, see Fig. 1b). The effective radius of the lowest primary exciton state on Ag(111) is small ($\sim 1.9$ \AA ~for e-h reduced mass $\mu=2.84$) which results in its weak overall dipolar coupling to the surface electronic response and thereby in a low probability of exciton dissociation via off-the-energy shell transitions into higher energy states. Therefore the duration of this primary exciton state is dominantly determined by the saturation of screening of the bare electron-SS hole potential. In the next stage the absorption of the second photon from either pulse brings the electron into an intermediate state in the Kepler quasicontinuum with a much larger effective excitonic radius. In this state with more spatially separated exciton constituents their monopole couplings  to the surface screening charge are more efficient in driving the exciton dissociation via energy conserving electron transitions from the Kepler-like into IP-states. This gives rise to very complex time dependence of the intermediate state polarization encompassing the excited  electron and SS-hole. Finally, in the third stage of perturbative photoemission the evolution of intermediate state polarization is probed by the absorption of the third photon which lifts the two-photon excited  electron into a final photoelectron state $|\phi_{f}\rangle$ of energy $E_{f}$ above the vacuum level. This makes possible to monitor the processes of transformation of initially coherent transient excitonic states$^{24}$ into the uncorrelated IP-electron and SS-hole band states$^{R3}$. Here it should also be noted that due to interactions of quasiparticles with the environment (heatbath) the individual SS-hole and IP-electron states are subject to decay and dephasing on the ultrashort time scale as described in Refs. 28 and R4-R6. Here the absorption edge singularities$^{R7}$, being a manifestation of the long time response to localized perturbations, are not expected to affect the excitonic spectrum on the discussed ultrashort time scale and for the present low effective masses of excited quasiparticles$^{R8}$.

To put the above described scenario of perturbative 3PP on the grounds of a tractable model for discussion of the experimental data we assume one-electron transitions induced by two identical delayed ultrashort pulses over a restricted set $\{|\phi_{1}(t)\rangle, |\phi_{Kep}(t)\rangle, |\phi_{IP}(t)\rangle\}$ of intermediate excitonic and image potential states. Their temporal variations, that means the evolution of the first (lowest) excitonic state $|\phi_{1}(t)\rangle$ and the Kepler-like states $|\phi_{Kep}(t)\rangle$, as well as the formation of IP-states  $|\phi_{IP}(t)\rangle$ are governed by the effective one-electron potentials $V^{exc}(t)$ and $V^{IP}(t)$ whose attenuation and formation, respectively, is modulated by the transient screening factor appropriate to Ag(111) surface (cf. Sec. 5 and Figs. 5 and 6 of Ref. 24). In accord with this scenario we consider stepwise 3PP from the SS-band to proceed via a sequence of one-electron transitions

\begin{center}
\bq
|\phi_{SS}\rangle \rightarrow |\phi_{1}(t')\rangle \rightarrow |\psi_{int}(t'')\rangle\rightarrow |\phi_{f}(t''')\rangle 
\label{eq:transitions}
\eq
\end{center}
where  $|\psi_{int}(t'')\rangle$ denotes the intermediate state in which the two-photon excited electron undergoes diabatic electronic transitions between the waning  Kepler-like and emergent IP-states. The required $|\psi_{int}(t'')\rangle$ is expressed as a linear combination of $\{|\phi_{Kep}(t'')\rangle,|\phi_{IP}(t'')\rangle\}$ with appropriatelly determined time dependent coefficients and phases$^{R9}$ driven by the attenuating $V^{exc}(t)$ and rising $V^{IP}(t)$. The SS-hole is assumed to retain its identity throughout the multiple excitation process (\ref{eq:transitions}) except for its dephasing and decay elaborated in Ref. 28.

A full solution to the thus defined  problem still poses a formidable task and further simplification of the model is necessary. Here we shall represent  $|\psi_{int}(t'')\rangle$ in the asymptotic form of a two-component Landau-Zener type of wave function$^{R9}$ composed of the unscreened excitonic $|\phi_{Kep}(t''\rightarrow 0)\rangle=|\phi_{Kep}\rangle$ and relaxed $|\phi_{IP}(t''\rightarrow\infty)\rangle=|\phi_{IP}\rangle$ diabatic states with the time dependent coefficients and phases induced by the effective excitonic and electron-image potentials. This rather restrictive approximation to otherwise very complex dynamics of intermediate electronic states facilitates the description of perturbative photoemission and yields the most salient features of the two-pulse correlation measurements for discussion of the experimental data. Taking into account that final state electrons are recorded in a detector which integrates the final state occupation over an interval that exceeds the time scale(s) of excitation and ultrafast screening processes, we define the 3PP two-pulse correlation in analogy to the 2PP case$^{12}$

\bq
{\rm Yield_{3PP}}(E_{f},\tau)=\int dt_{f} \langle \Psi(t_{f},\tau)|\hat{n}_{f}|\Psi(t_{f},\tau)\rangle
\label{eq:3PYield}
\eq
where $\Psi(t_{f},\tau)$ is the system wavefunction at time $t_{f}$ calculated perturbatively in the length gauge to third order in powers of the interaction of the system with total laser electric field ${\cal E}(t,\tau)={\cal E}_{1}(t)+{\cal E}_{2}(t-\tau)$, and $\hat{n}_{f}$ is the electron occupation number operator for the state $|\phi_{f}\rangle$.

Polarization involving transient excitonic intermediate states gives rise to specific features in the interferometric yield defined  by Eq.  (\ref{eq:3PYield}) that distinguish it from the two-pulse third-order interferometric autocorrelation (IAC) function$^{R10}$

\bq
G_{3}(\tau)=\int\left|[{\cal E}_{1}(t)+{\cal E}_{2}(t-\tau)]^{3}\right|^2 dt
\label{eq:IAC}
\eq
which is used as a standard reference in the discussions of interferometric correlations. Hence, in the following we shall compare the simulations of 3PP correlation yield with the IAC function of the corresponding pulses. Here we note that for identical pulses ${\cal E}_{1}(t)={\cal E}_{2}(t)$ used in the described experiments one obtains $G_{3}(-\tau)=G_{3}(\tau)$ which also holds for ${\rm Yield_{3PP}}(E_{f},\tau)$. This property will be exploited below in the presentation of the results of simulation.

As an illustration of the features predicted by the simple TE model we use Eq. (\ref{eq:3PYield}) to simulate normal interferometric 3PP yield from the SS-band on Ag(111) for the specific case of laser excitation energy $\hbar\omega_{laser}=\hbar\omega_{res}=(E_{IP}-E_{SS})/2=1.93$ eV at which  the one photon-induced transitions $|\phi_{SS}\rangle \rightarrow |\phi_{1}\rangle$ are offresonant, whereas the two photon-induced transitions into the intermediate state $|\psi_{int}(t'')\rangle$ are resonant with the   bottom of the fully developed IP-band. These conditions give  rise to the spectrum depicted in Fig. 2d, but also to the two-photon correlation in Fig. 4e when the effects of the local field are dominant, and which are important for revealing the TE features in 3PP spectra from Ag(111) SS-band. Due to the fact that the unscreened Kepler-like states make a quasicontinuum below the upper edge of the surface projected bulk {\it s,p}-band gap, there  always exists a Kepler-like level almost degenerate with the relaxed (screened) IP-level so that electron transitions into $|\psi_{int}(t'')\rangle$ with $\epsilon_{Kep}\approx E_{IP}$ are least affected by the time-energy uncertainity during the evolution interval. This maximum resonant regime minimizes the transients and strongly enhances the amplitudes of two-photon induced polarization probed by the third photon in yet another resonant process. 
The matrix elements (oscillator strengths) of the transitions from symbolic relation (\ref{eq:transitions}) factorize out of the triple intermediate time integrals constituting  $\Psi(t_{f},\tau)$, likewise in the case of 2PP amplitudes$^{R11,R12}$. Assuming them constant in the considered narrow energy intervals$^{R12}$ they appear as overall scaling factors on the right-hand-side of Eq. (\ref{eq:3PYield}). The results of model simulations for normal 3PP yield induced by two pulses of resonant energy $\hbar\omega_{res}$ and full width at half maximum FWHM=15 fs  are presented in the upper panel of Fig. S2 together with the IAC function  of the applied electric fields defined in Eq. (\ref{eq:IAC}). 

The thus calculated 3PP yield follows the trends observed experimentally in two-pulse correlation measurements of 3PP from the SS-bands on Ag(111), see Fig. 4e of the main text. First, the dominant fringe structure apears at multiples of $1\times \hbar\omega_{laser}$ (which is here the resonant frequency), although the multiples of $2\times \hbar\omega_{laser}$, and to a much lesser extent of $3\times \hbar\omega_{laser}$, are discernible (cf. Fig. S1). This is indicative of stepwise excitation illustrated symbolically by relation (\ref{eq:transitions}). Second, the prounounced fringe structure lasts long beyond the direct overlap of the two pulses, i.e. much longer than the interference in their autocorrelation. This indicates relatively long lifetime of TE and IP intermediate states in the sequence shown in expression (\ref{eq:transitions}), and the evolution (dephasing) of the former into the latter. It is also gratifying that the maximum amplitude ratio or saturation scaling

\bq
\frac{{\rm Yield_{3PP}}(E_{f},\tau=0)}{{\rm Yield_{3PP}}(E_{f},\tau\rightarrow\infty)}
\sim  \frac{G_{3}(\tau=0)}{G_{3}(\tau\rightarrow\infty)}=32
\label{eq:ratio}
\eq
which indicates and controls the consistency of simulations of multi-photon photoemission yields$^{R11}$, is here fulfilled with high accuracy.

\begin{center}
\rotatebox{0}{\epsfxsize=13cm \epsffile{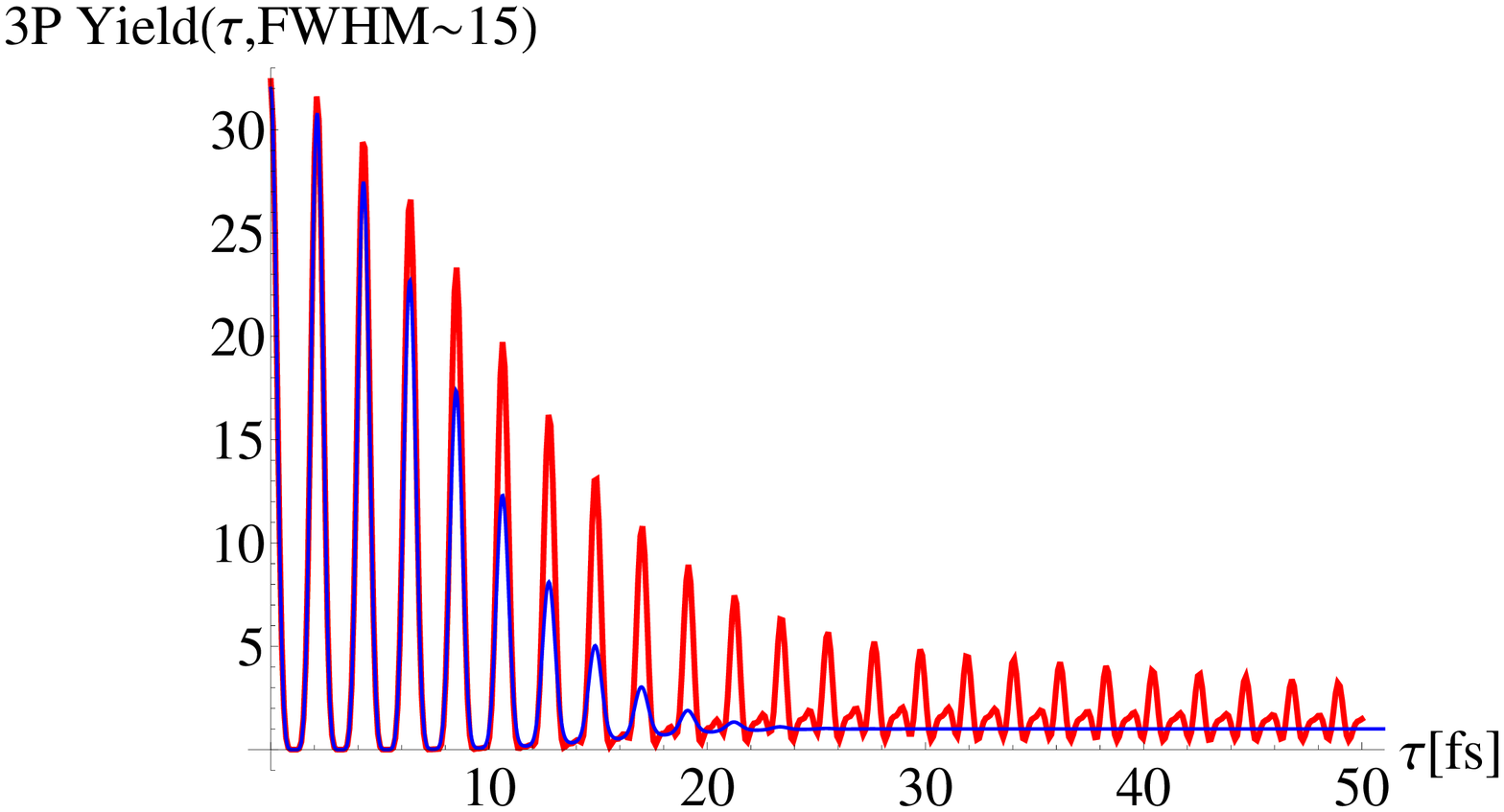} } %8.5cm
\end{center}
\begin{center}
\rotatebox{0}{\epsfxsize=13cm \epsffile{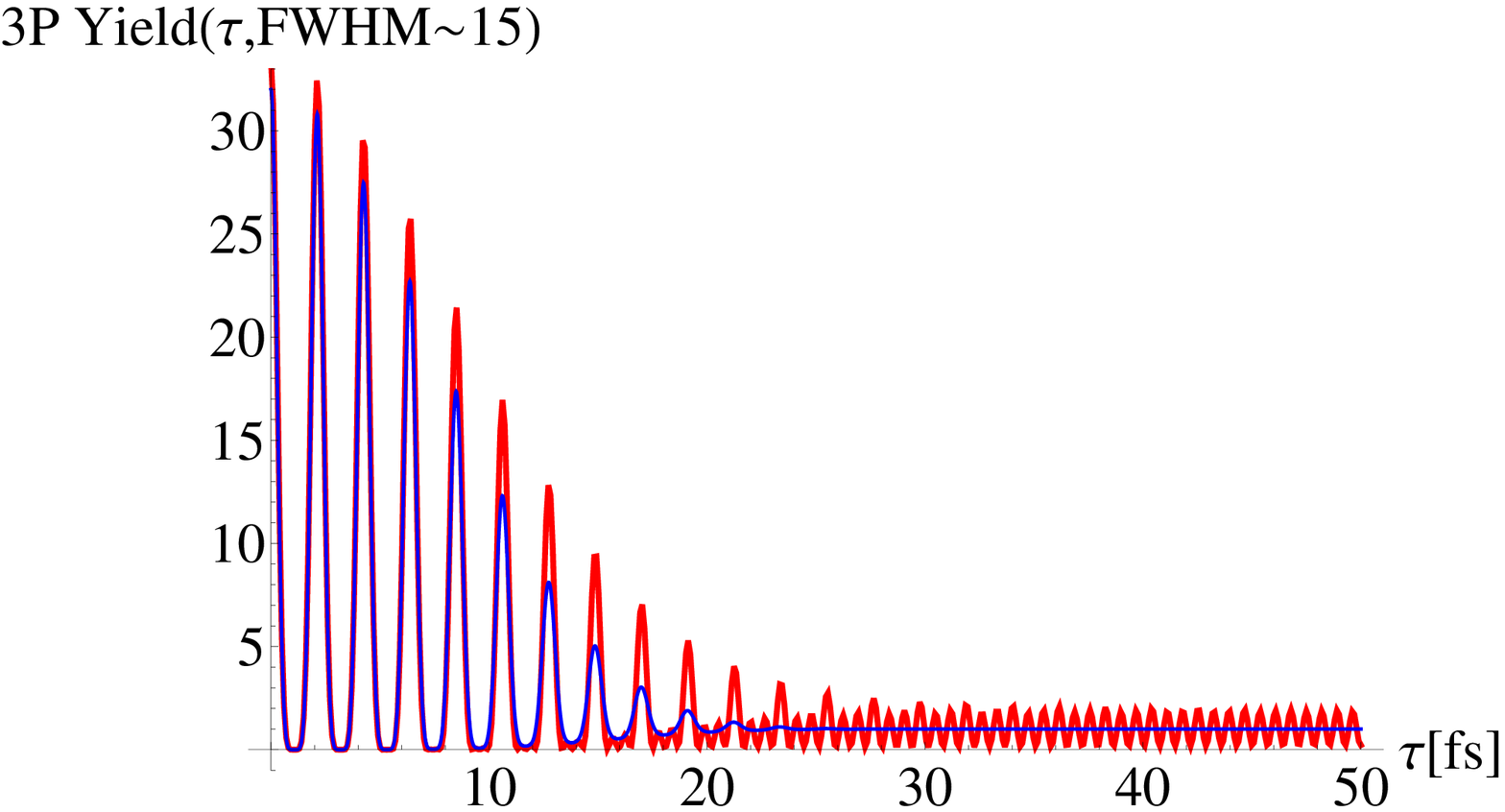} } %8.5cm
\end{center}
\noindent {\small {\bf Figure S2.} {\it Upper panel:} Simulation of interferometric 3PP yield from Ag(111) surface (red curve) for resonant two-pulse excitation with $\hbar\omega_{laser}=1.93$ eV and the corresponding third order pulse IAC (blue curve) based on expressions (\ref{eq:3PYield}) and (\ref{eq:IAC}), respectively, and plotted as the functions of pulse delay $\tau$. Both spectra are scaled to their asymptotic or saturation values for $\tau\rightarrow\infty$. The full width at half-maximum  of applied pulses is FWHM=15 fs. 
 {\it Lower panel:} Simulation of a restricted interferometric 3PP yield from Ag(111) surface (red) obtained by including only the TE states  in the intermediate state $|\psi_{int}(t'')\rangle$ of the sequence in expression (\ref{eq:transitions}). Pulse parameters $\hbar\omega_{laser}$ and FWHM, IAC (blue curve), and spectral scaling same as in the upper panel.}
\vskip 0.3 cm
To get further insight into the role of interplay of TE and emergent IP-states in building the photo-induced intermediate state polarization on the time scale of screening of the bare excitonic potential we simulate a restricted form of expression (\ref{eq:3PYield}) by allowing only the evolving TE states to partake in the intermediate state $|\psi_{int}(t'')\rangle$, but not the developing IP-states which are not instantly available. The result is shown in the lower panel of Fig. S2. Comparison of simulations of the 3PP two pulse correlations shown in the two panels indicates that  the polarization proceeding via TE states gives a dominant contribution to the 3PP yield for the delays  shorter than the FWHM of applied pulses. The major difference between the spectra appears outside the temporal overlap of the two pulses, $\tau\geq 15$ fs, which also coincides with asymptotic saturation of the image potential on Ag(111) surface and formation of IP-states. It is seen that if the IP-states are not available in two-photon induced transitions, the 3PP signal is restricted to the pulse overlap during exciton lifetime and, moreover, is dominated by the $3\times\hbar\omega_{res}$ oscillations after the pulse overlap has died out. These higher order fringes are  reminiscent of the ones observed in two-pulse  correlations in 2PP from SS-band on Cu(111) surface under nonresonant conditions when the IP-state does not participate in energy conserving excitations$^{13}$. By contrast, if the IP-states are not excluded from the intermediate resonant state the oscillations at $1\times\hbar\omega_{res}$ are still dominant on the time scale of SS-hole and IP-electron dephasing (upper panel). In other words, the presence of both the waning TE and emergent IP-states is needed in  $|\psi_{int}(t'')\rangle$ in order to simulate two-pulse correlations with the features closer to the observed ones.  This reinforces the importance of the interplay of TE and IP-states in multi-photon photoemission from Ag(111) surface state band and provides a framework for explanation of the high intensity nondispersive feature in 3PP spectra discussed in the main text.

\noindent {\bf Supplementary references:}
\vskip 3mm

\small{

\noindent $^{R1}$\hspace{2mm} \parbox[t]{16cm}{Born, M. \& Fock, V. Beweis des Adiabatensatzes.  {\it Z. Physik  A} {\bf 51}, 165 (1928).}

\noindent $^{R2}$\hspace{2mm} \parbox[t]{16cm} {L.I. Schiff, Quantum Mechanics. {\it McGraw-Hill Book Company, New York, 1968}, \S  Adiabatic approximation, p. 289.}

\noindent $^{R3}$\hspace{2mm} \parbox[t]{16cm}{El-Shaer, F. \& Gumhalter, B. Entangled and disentangled decoherence of intermediate electron-hole pairs in two-photon photoemission from surface bands: Beyond the adiabatic approximation. {\it Phys. Rev. Lett.}  {\bf 93}, 236804 (2004).}

\noindent $^{R4}$\hspace{2mm} \parbox[t]{16cm}{Lazi\'{c}, P., Silkin, V.M., Chulkov, E.V., Echenique, P.M. \& Gumhalter, B. Extreme Ultrafast Dynamics of Quasiparticles Excited in Surface Electronic Bands. {\it Phys. Rev. Lett.} {\bf 97}, 086801 (2006).}

\noindent $^{R5}$\hspace{2mm} \parbox[t]{16cm}{Lazi\'{c}, P., Silkin, V.M., Chulkov, E.V., Echenique, P.M. \& Gumhalter, B. Ultrafast dynamics and decoherence of quasiparticles in surface bands: Preasymptotic decay and dephasing of quasiparticle states. {\it Phys. Rev. B} {\bf 76}, 045420 (2007).}

\noindent $^{R6}$\hspace{2mm} \parbox[t]{16cm}{Lazi\'{c}, P., Aumiler, D. \& Gumhalter, B. Nonadiabatic quasiparticle dynamics in time resolved electron spectroscopies of surface bands. {\it Surf. Sci.} {\bf 603},1571 (2009).}

\noindent $^{R7}$\hspace{2mm} \parbox[t]{16cm}{Mahan, G.D., Collective excitations in x-ray spectra of metals. {\it Phys. Rev. B} {\bf 11}, 4814 (1975); and references therein.}  

\noindent $^{R8}$\hspace{2mm} \parbox[t]{16cm}{Gavoret, J., Nozi\`{e}res, P., Roulet, B., and Combescot, M., Optical absorption in degenerate semiconductors. {\it J. Phys. (Paris)} {\bf 30}, 987 (1969).} 

\noindent $^{R9}$\hspace{2mm} \parbox[t]{16cm}{Zener, C. Nonadiabatic crossing of energy levels. {\it Proc. Roy. Soc. A} {\bf 1R7}, 696 (1932).}

\noindent $^{R10}$\hspace{2mm} \parbox[t]{16cm}{Diels, J.-C. \& Rudolph W. Ultrashort Laser Pulse Phenomena, {\it (Elsevier, 2006)}, Ch. 9.2.} 

\noindent $^{R11}$\hspace{2mm} \parbox[t]{16cm}{Timm, C. \& Bennemann K.H. Response theory for time-resolved second-harmonic generation and two-photon photoemission. {\it J. Phys.: Condens. Matter} {\bf 16}, 661 (2004).} 

\noindent $^{R12}$\hspace{2mm} \parbox[t]{16cm}{Ueba, H. \& Gumhalter, B. Theory of two-photon photoemission spectroscopy of surfaces. {\it Prog. Surf. Sci.} {\bf 82}, 193 (2007).} 

}

\end{document}